# NUMBER COUNTS, CONFUSION, MAPPING ISSUES AND SKY COVERAGE ANALYSIS FOR RADIO CONTINUUM SURVEYS THROUGH EMU EARLY SCIENCE, EMU-ASKAP, AND WODAN ESPECIALLY FOR COSMOLOGY SCIENCE GOALS.

**Syed Faisal ur Rahman[a]*, Muhammad Jawed Iqbal[a]**

[a]*Institute of Space Science and Technology (ISST), University of Karachi, 75270,Karachi,Pakistan,*

*e-mail(*):* faisalrahman36@hotmail.com

**Abstract**: In this study, we discuss the challenges radio astronomers face while observing radio continuum sources. We will discuss issues related to rms noise, confusion, position accuracy, shot noise and how these issues can affect observation results, data analysis and the science goals we are trying to achieve. We will mainly focus on the Evolutionary Map of the Universe (EMU-ASKAP) sky survey, EMU Early science survey and Westerbork Observations of the Deep APERTIF Northern sky (WODAN), for our study. The study will also be useful for future surveys like with possible continuum surveys through MeerKAT (e.g. MIGHTEE) and SKA-1. The late time Integrated Sachs-Wolfe (ISW) effect detection is one of the major areas of research related to dark energy cosmology. We will particularly discuss how technical, data analysis and mapping issues, affect galaxy over/under density dependent science goals like the detection of the late time Integrated Sachs-Wolfe (ISW) effect through wide-field radio continuum surveys.

## INTRODUCTION

With advances in antenna technologies, imaging techniques, and powerful computing machines, it has become possible to look deeper into our sky, a lot more than was imaginable few decades ago (see: [1], [2], [3], [4], [5], [6], [7], [8], [9], [10], [11], [12], [13], [14], [15] and [16]). These developments provided hope for understanding some of the most important questions about our universe and its cosmic history especially in settling questions related to the correct values for dark energy density parameters and expansion rate( read ref: [8], [9], [17], [18], [19] and [20]). We now have the capability to observe millions of galaxies with highly sensitive telescopes ([8], [21], [22]).

Keeping in mind these advances and the future course of radio astronomy, especially in relation to the wide field radio continuum surveys for science goals, such as the Integrated Sachs-Wolfe effect ([23], [24], [25], [26], [27], [28]), galaxy auto-correlations ([8],[29]), cosmic magnification ([30], [31], [32], [33], [34]) and others, we need to pay extra attention to the technical parameters and how they affect the completion of our science goals.

This study will first focus on the technical aspects like confusion, position accuracy, and shot-noise. We will also study how the power law distribution of sources can help us in estimating these technical measures. After that, we will move towards discussing some issues related to the data analysis and mapping issues of the wide field continuum surveys.

We will particularly focus on the late time Integrated Sachs-Wolfe effect as discussed in ([35], [36], [37], [38]) and see how number counts and sky coverage, along with the redshift range and the maximum multipole range, affect the ISW signal to noise ratio analysis for surveys like the Evolutionary Map of the Universe (EMU-ASKAP), EMU Early Science Cosmology and WODAN surveys using simulated data ([8], [21], [35],[36], [39]). While there were previous studies ([4], [9], [18]) which dealt with the cosmology from upcoming radio continuum surveys, there is still a gap in the literature where key technical and mapping issues are discussed in relation to their effects on major science goals especially when it comes to looking at the combine effects of shot-noise, confusion, redshift range, sky coverage and maximum multipole range.

## EMU-ASKAP

Evolutionary Map of the Universe (EMU) will be the most sensitive large scale radio continuum survey of its time before Square Kilometer Array’s first phase (SKA-1) is launched ([35], [36]). It will cover around 75% or around ($3\pi$ steradians) of the sky. It will include a full coverage of the southern hemisphere and the equatorial region, right to +30 degrees declination. The sensitivity of the survey is expected to be around 10 micro Jy/beam. The results presented here will mostly be related to the EMU-ASKAP Cosmology Science goal. However, the discussions will also be useful for other areas as well, which rely on galaxy counts ([35],[36]).EMU-ASKAP will have a resolution of around 10 arcseconds ([35],[36]).

## EMU-EARLY SCIENCE COSMOLOGY

Before the start of the EMU-ASKAP full survey, there will be an Early Science survey ([35], [36]) which will use ASKAP-12 configuration. The survey results will not only be used to analyze the technical aspects of the telescope but it will also help in developing data analysis pipelines and conduct some interesting science based on EMU Early Science Survey alone.

## WODAN

As a northern sky survey, with similar sensitivity as of EMU-ASKAP and coverage of >+30 degrees declination coverage, Westerbork Observations of the Deep APERTIF Northern sky (WODAN) will be well suited to complement the full EMU-ASKAP survey ([9], [21]). WODAN is going to cover about 25% (1π steradians) of the sky on the northern hemisphere ([9], [21]). It will have an expected resolution of about 15 arcseconds.

## CONFUSION AND POSITION ACCURACY OF EMU_ASKAP EARLY SCIENCE, FULL EMU-ASKAP, AND WODAN

To ensure reliable individual detection of sources in a radio continuum survey, we need to keep in check the rms confusion limit. The rms confusion limit determines if a source can be resolved independently with no position or count uncertainty. Confusion analysis is not just important for galaxy count maps but also for diffuse emissions mapping for science goals such as the EMU-ASKAP Synchrotron Cosmic Web [38].

In order to calculate the rms confusion limit, we can fit the power law curve for the differential source count of the continuum survey ([1], [2], [3], [4], [5], [6], [8], [9], [40], [41], and [42]). We can obtain the differential source count power law probability distribution ([9], [42]) in $Jy^{-1}\ Sr^{-1}$as:

$$n(S) = \frac{dN}{dS} = kS^{-\gamma} \tag{1}$$

Here, n(S) is the differential source count tailed probability distribution and S is the flux in Jy. We can see the lower bound value for Smin>0 as it will be undefined for Smin=0Jy/beam which is also not a realistic value. We first use SKADS (S3- SEX) ([39]) to get the differential source counts for 1.4 GHz sky with a flux density range of 50 micro Jy/beam < S<570 micro Jy/beam with a redshift range of 0<z<5.8, and obtain k=57.24 and γ=2.18 (51). Condon 2007, uses flux range of 1 micro Jy < S <100 micro Jy and estimated, k=1000 and γ =1.9. Similarly, Kellermann 2000 measured k=8.23 and γ =2.4 and Mitchell & Condon 1985 estimated k=57 and γ=2.2.

We also estimated power law distributions using SKADS and obtained results as given in table (1)

| | S_min (uJy/beam) | S_max (uJy/beam) | k | γ |
|---|---|---|---|---|
| Sample-1 | 50 | 100 | 60.477 | 2.176 |
| Sample-2 | 100 | 200 | 24.34 | 2.274 |
| Sample-3 | 200 | 350 | 55.137 | 2.177 |
| Sample-4 | 350 | 570 | 200.55 | 2.017 |
| Sample-5 | 800 | 870 | 3.193 | 2.608 |

**Table 1-k and gamma values for different power law distributions from samples generated using Square Kilometer Array Design Studies (SKADS) database. We can observe the dependency of power-law distribution on the selection of flux ranges.**

As we can see from the table (1) that k and γ, greatly depend on the range of minimum and maximum flux densities. This thing is particularly important to keep in mind while working on the science goals like cosmic magnification or ISW effect where flux cuts are applied to the catalog data based on technical or science goal requirements.

Figure (1) shows the flux normalized plots for the probability distributions obtained from SKADS and their comparison with previous studies ([1], [2], [3], [4], [5], [6], [8], [9], [40], and [41])

In statistical analysis, it is useful to observe the sample value probability via the complementary cumulative probability distribution function (CCDF). This will indicate how often we can expect a source detection or presence in a sample population using ([42]):

$$\Pr(S>Smin) = \left(\frac{S}{Smin}\right)^{-\gamma+1}$$

(2)

In figure (2), we can observe CCDF for the differential source count samples. This will also give us some scale invariance perspective, especially in the case of the differential source counts where the size of flux bins can also affect the distribution shape.

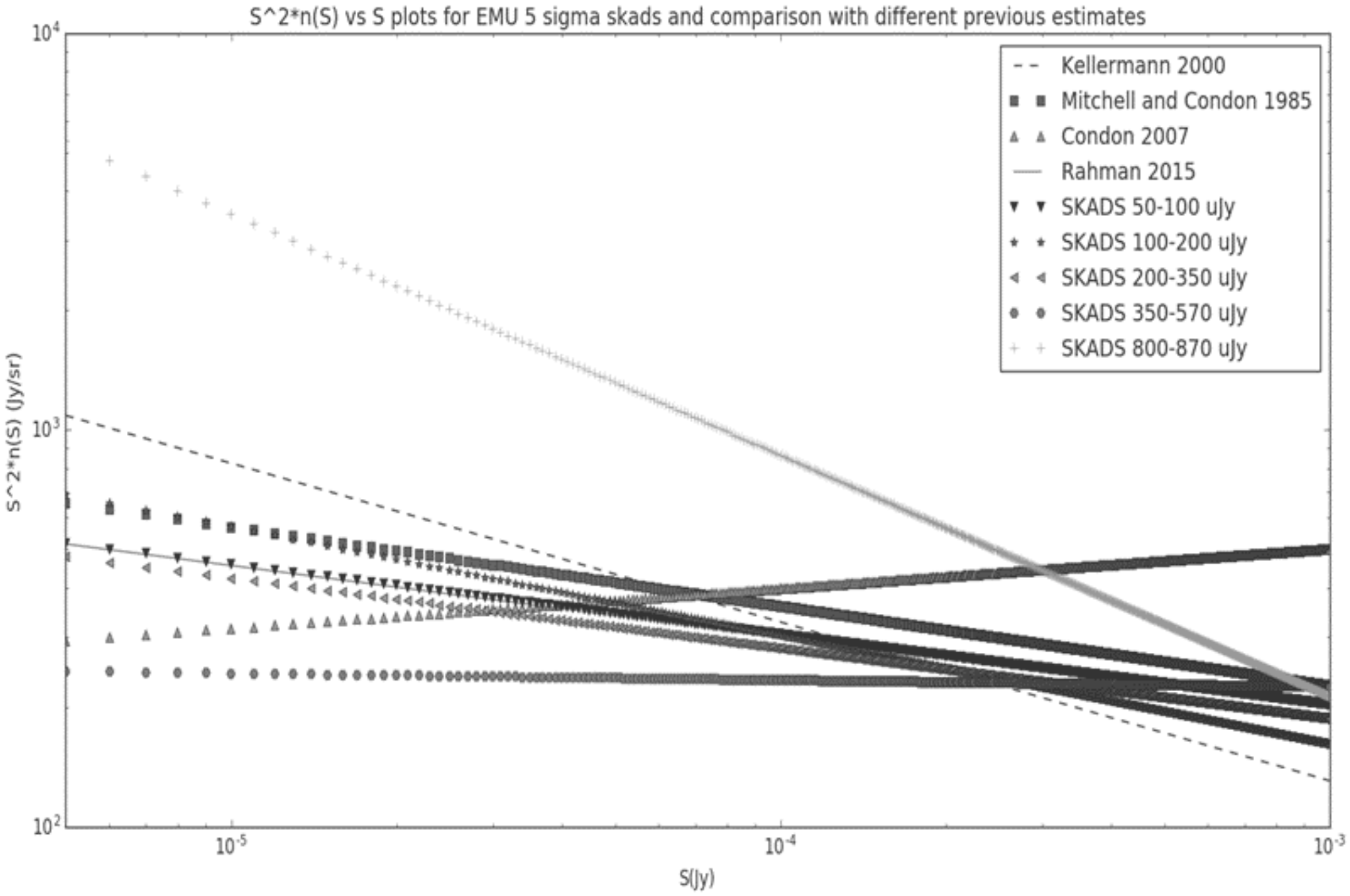

**Figure 1-Differential power law fits plots from the probability distributions discussed in Kellerman 2000 (dashed lines), Mitchell and Condon 1985 (squares), Condon 2007 (up triangles) and Rahman 2015 (line). We have also shown estimates from SKADS database for sources with fluxes 50-100 μJy/beam (down triangle), 100-200 μJy/beam (stars), 200-350 μJy/beam (left triangle), 350-570 μJy/beam (Hexagon), and 800-870 μJy/beam (plus signs). For Rahman 2015 and SKADS results, we used $\Delta S$=5μJy/beam.**

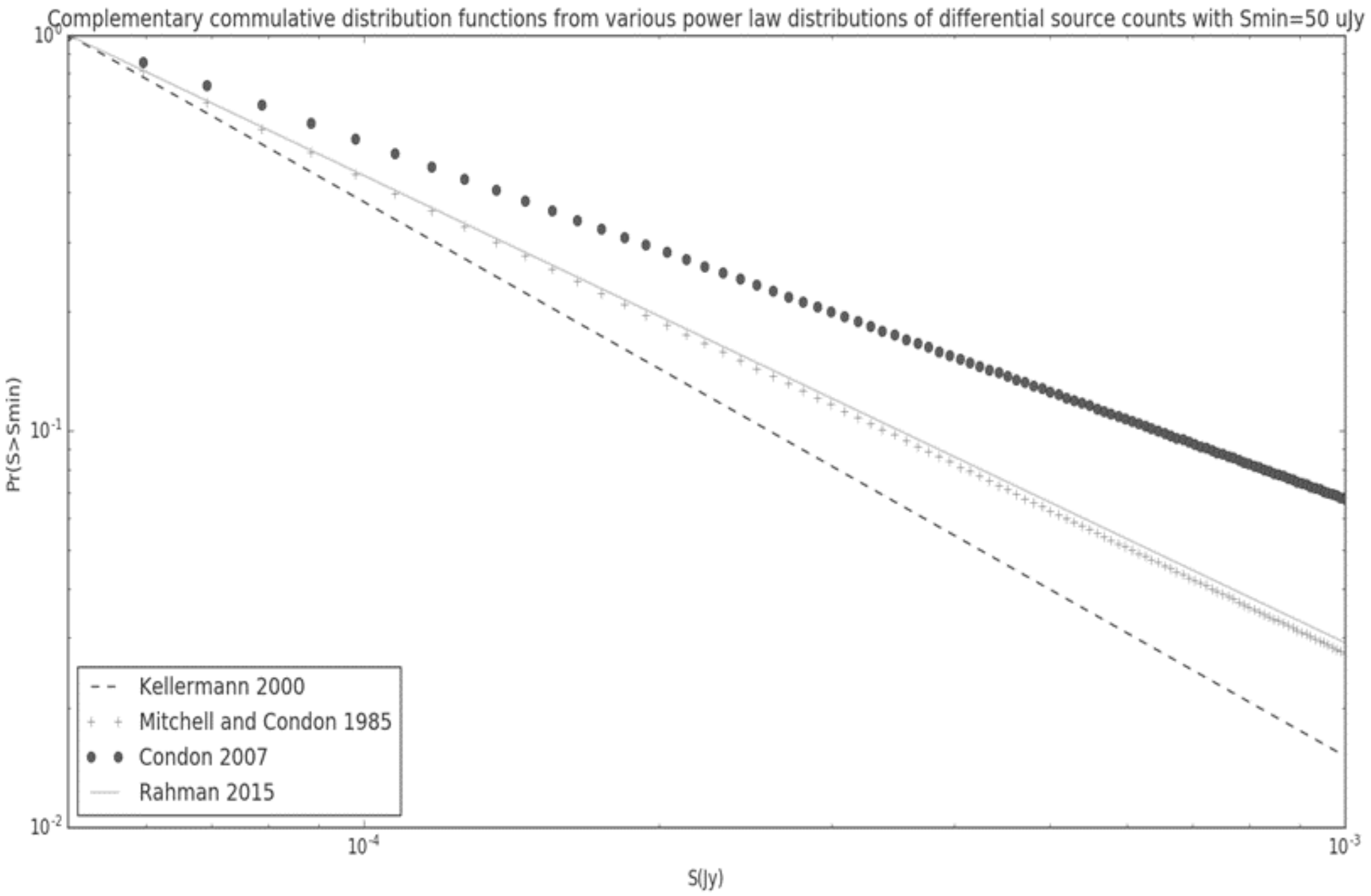

**Figure 2- Cumulative probability distribution function plots for Kellermann 2000, Mitchell and Condon 1985, Condon 2007 and Rahman 2015.**

In order to estimate the rms confusion, we will first need to calculate the beam solid angle. Considering a target resolution of around 30" beam for ASKAP-12 like configurations, we can get the beam solid angle as ([1], [4], and [9]):

$$\Omega_b = \frac{\pi\theta^2}{4ln2}$$

Which is around 2.397 e-08 giving effective solid angle $\Omega_e = \Omega_b/\gamma$-1 of approx. 2.03 e-08. Effective solid angle caters the effects of side lobes of the PSF due to the decline in gamma

We can calculate the confusion as ([1], [4], and [9]):

$$\sigma_c = \left(\frac{q^{3-\gamma}}{3-\gamma}\right)^{\frac{1}{\gamma-1}} (k\Omega e)^{\frac{1}{\gamma-1}} \quad (3)$$

Here q is taken as 5 for some intensity $I_{max}=q\sigma_c$ ([1], [4]) in order to ensure reliable detection of sources. The intensity( I) comes from the probability distribution p(I) which indicates the probability of any arbitrary point on a noiseless astronomical image having an intensity I. In case of the rms confusion, it will be in Jy/beam. This will lead us to the rms confusion of about 33.87 microJy/beam for k=57.24 and γ=2.18. Table 2 provides estimates for some more related studies ([1], [2], [3], [4], [5], [6], [8], [9], [40], and [41])

| k | γ | The rms confusion (micro Jy/beam) |
|---|---|---|
| 57.24 | 2.18 | 33.8 |
| 1000 | 1.9 | 39.3 |
| 8.23 | 2.4 | 41.4 |
| 57 | 2.2 | 39.7 |

**Table 2-The rms confusion for different power law distributions given in Rahman 2015, Condon 2007, Kellermann 2000 and Mitchell and Condon 1985. The estimates are for a 30" beam.**

ASKAP-12 will have a baseline of at least 2.3 km. On the other hand, full ASKAP, WODAN and future SKA-1 configurations will enjoy a much higher range. This will provide a much better resolution and so their rms confusion range will also be affected by this. In figure 3, we have calculated the rms confusion for some possible resolutions with k=57.24 and γ=2.18.

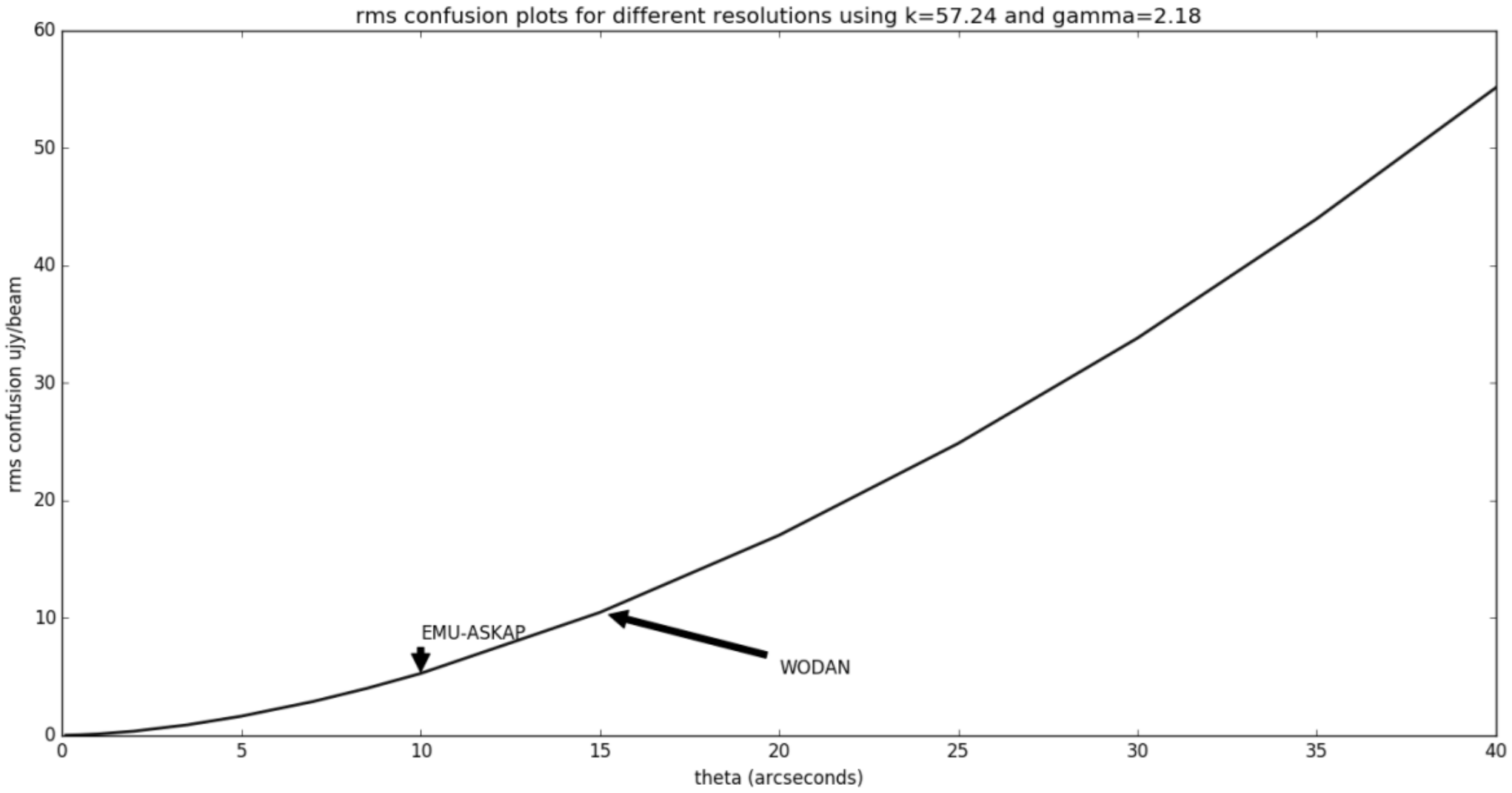

**Figure 3-The rms confusion for different resolutions with k=57.24 and γ=2.18. EMU-ASKAP is expected to have 10” resolution, WODAN will have around 15” and EMU-Early Science will be expected to have around 25-35” resolution ranges. Resolutions below 10” are for future surveys like, VLASS, MIGHTEE tier 2 & 3 and surveys from the SKA.**

We can obtain total rms $\sigma_{tot}$ as ([1], [4]):

$$\sigma^2_{tot} = \sigma^2_n + \sigma^2_c \quad (4)$$

We can calculate the rms noise, $\sigma_n$ as ([12]):

$$\sigma_n = \frac{2kTsys}{Aeff\sqrt{N(N-1)}\sqrt{\Delta\nu t}} \text{ (Jy/beam)} \quad (5)$$

Where k=1380 Jy $m^2$ $K^{-1}$ is the Boltzmann constant, Tsys is the average system temperature, Δv is bandwidth, t is the survey integration time, N is the number of antennas in the array, $A_{eff}$ is the effective collection area which can be calculated using the relation:

$A_{eff}=\eta_a A$

Where $\eta_a$ is the aperture efficiency and $A$ is the area of the antenna.

We can calculate the position accuracy as ([1], [4]):

$$\sigma_p \approx \frac{\sigma_{tot}\theta}{2S} \quad (6)$$

EMU-ASKAP full has a 1-sigma rms noise sensitivity of 10 µJy/beam and a 10-arcsecond resolution. Which gives $\sigma_c$=5.26 µJy/beam and $\sigma_{tot}\approx$ 11.299 µJy/beam. This will give us $\sigma_p\approx$1 arcsec at S≈56.5 µJy/beam.

WODAN has a 1-sigma rms noise sensitivity of 10 µJy/beam and a 15-arcsecond resolution. Which gives $\sigma_c$=10.453 µJy/beam and $\sigma_{tot}\approx$ 14.466 µJy. This will give us $\sigma_p\approx$1 arcsec at S≈ 108.495 µJy/beam.

In case of the upcoming EMU-Early Science configuration, with 30" beam, for $\sigma_n$=100µJy/beam and $\sigma_c$=33.8 µJy, we get $\sigma_{tot}\approx$ 105.5 µJy/beam. For a target position accuracy of 1 arcsecond, we can go for S ≈1582 µJy/beam and for 10-arcsecond position accuracy at S≈158.2 µJy/beam.

The MIGHTEE survey configuration ([43],[44],[45], and [46]) is aimed at a 20 square degrees survey with ~6 arc-seconds resolution and 1 µJy/beam sensitivity. This will give an rms confusion of ~2.212 µJy/beam and σtot≈ 2.4275 µJy/beam. This provides us $\sigma_p\approx$1 arcsec at S≈7.283 µJy/beam.

For the proposed MIGHTEE ultra-deep, will have an expected sensitivity of~0.2µJy/beam ([30, 31]). Using a resolution of ~3.5 arcsec, we can expect $\sigma_{tot}\approx$ 0.9 µJy/beam. This gives us $\sigma_p\approx$1 arcsec at S≈1.58 µJy/beam.

The proposed MIGHTEE surveys go much deeper than EMU-ASKAP full, WODAN or EMU-Early Science but their limited sky coverage restricts their ability to pursue science goals like galaxy-autocorrelations or ISW effect detection. However, they will be able to serve the goals like studying cluster evolution, star formation, and lensing and rotation measures. Another survey from National Radio Astronomy Observatory (NRAO), The Very Large Array Sky Survey (VLASS) , is expected to have higher resolution (~2.5 arcsec) and better position accuracy (0.2 arcsec) than EMU or WODAN but will be of lesser sensitivity ~69 micro Jy/beam. VLASS will also be useful as a companion survey for EMU because of its large coverage of Northern hemisphere (33885 square degrees) ([47], [48]).

## SOURCE COUNTS

In figure (4), we have shown theoretical estimates for the number of sources per steradian for various redshift ranges using ([39], [45]).

Theoretically, we can obtain integral source counts from differential count power law distributions as:

$$N(S>S_{min})=\int n(S)dS \tag{7}$$

However, in order to get results for a realistic survey, we cannot use indefinite integrals and we need to apply certain flux limits. We can restrict Smax=1Jy/beam and calculate the source count per steradian as shown in the figure (5) by using equation (7).

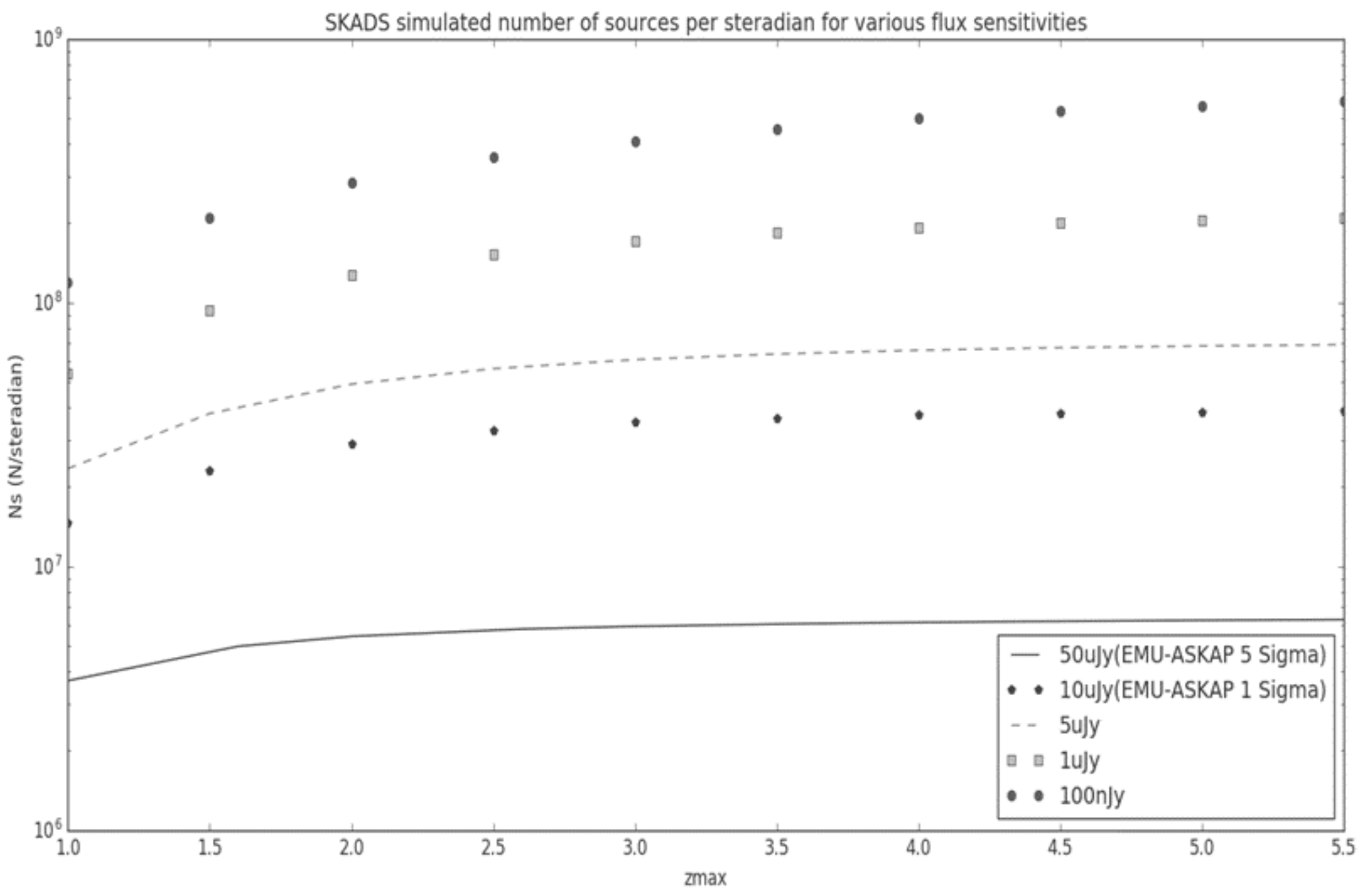

**Figure 4- Theoretically expected number of sources per steradian based on the maximum redshift range using (Jarvis et al. 2015; Rahman 2016; Rahman 2015) results. The vertical axis is log scaled to clearly see the differences in estimated source counts.**

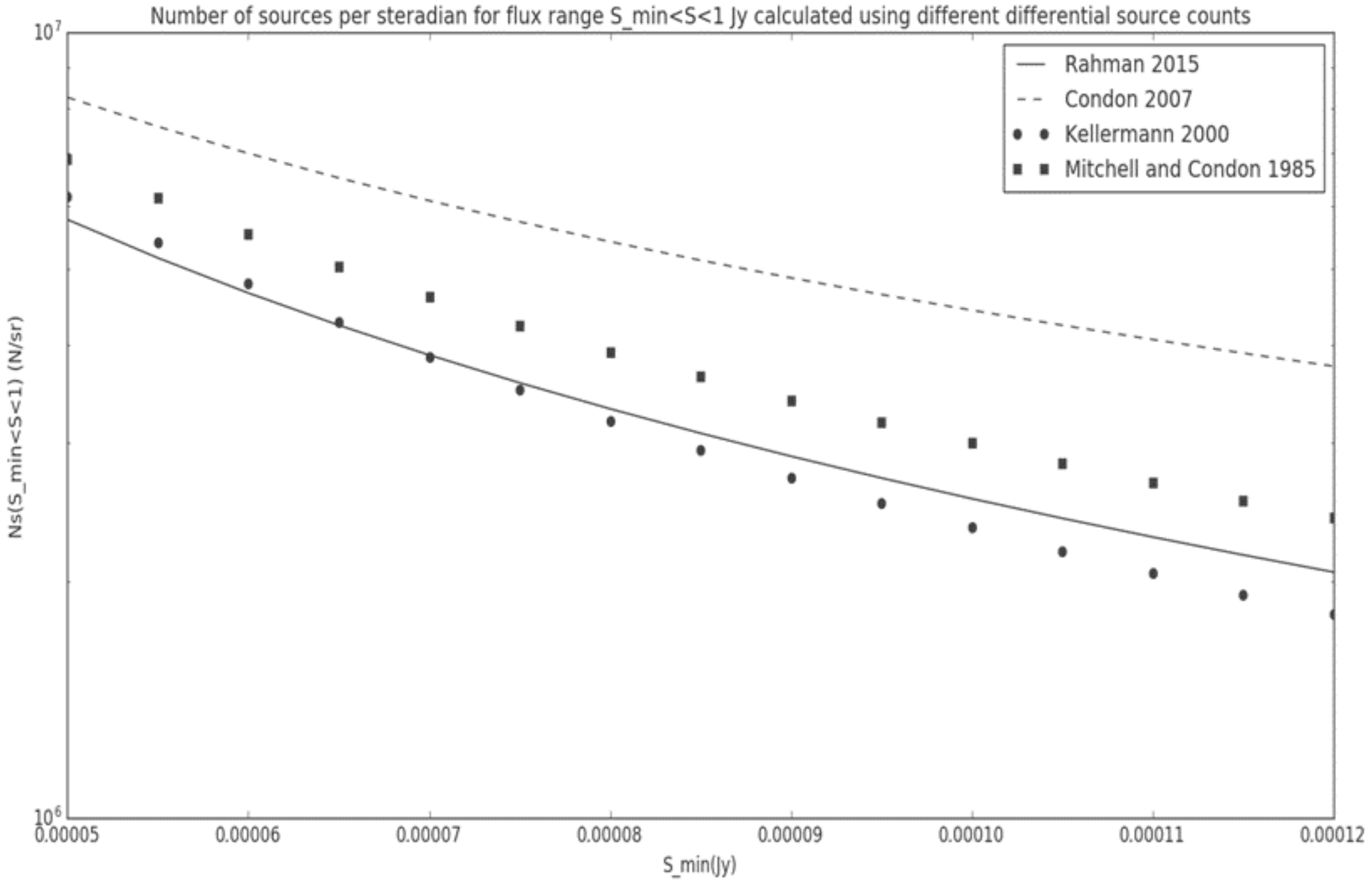

**Figure 5-Integral source counts using the power law distributions given in Rahman 2015, Condon 2007, Kellermann 2000 and Mitchell and Condon 1985. Here we restricted Smax=1Jy/beam. The vertical axis is log scaled to clearly see the differences in estimated source counts.**

In figure (5), we can also observe a downward slope of the integral source counts with respect to S_min. This can be quantified using the relation:

$$N(S)=CS^{-\alpha}$$

From equations (1) and (7), one may consider that we can simply perform an indefinite integral to obtain α from ϒ. However, this approach will lead to wrong slopes for actual or simulated counts. Using S_min range 50μJy/beam <S_min<120μJy/beam with a 5μJy / beam difference, we obtained power law curves for N (S>S_min) with Rahman 2015 and Condon 2007 distributions. We used S_max=1Jy/beam for our analysis.

| | Rahman 2015, using source count samples | Condon 2007, using source count samples | Rahman 2015, using the indefinite integral of n(S) | Condon 2007, using the indefinite integral of n(S) |
|---|---|---|---|---|
| α | 1.18002 | 1.11088 | 1.18 | 0.9 |
| C | 48.49996 | 2412.85 | -48.5085 | -1111.11 |

**Table 3--Power law distribution parameters for N(S>S_min) with S_max=1Jy/beam for n(S) distributions given in Rahman 2015 and Condon 2007. We can also see their difference with indefinite integral results of n(S).**

The measurement of α is important to study science goals like magnification bias in both galaxy surveys and cosmic microwave background studies ([37], [50], [51], [52]). Table 3 provides a comparison of estimates from Condon 2007 and SKADS sources discussed in Rahman 2015. EMU-ASKAP is expected to detect over 70 million sources and VLASS which will cover slightly more area than EMU with higher resolution will cover roughly 9.7 million sources because of lower sensitivity.

## SHOT-NOISE

In order to perform statistical error analysis ([51], [52], [53], [54], [55], and [56]) for galaxy continuum surveys, the shot-noise measurements play an important part. Shot noise estimates or measurements are required to calculate the signal to noise ratios, measure error bars, and obtain correct covariance matrices, especially in relation to the theoretical or observed ‘Cl’ values obtained during cross or autocorrelation studies. We can define shot-noise as ([1], [4])

$$Shot\ Noise = \frac{\Delta\Omega}{N} \qquad (8)$$

Where $\Delta\Omega$ =observed area of the survey in steradian and N=number of sources observed in the total survey area. Shot-noise can be calculated from the number count per steradian (Ns), by using the simple relation:

Shot-Noise=1/Ns

Where Ns is the number of sources per steradian.

Figure (6) shows shot-noise estimates over different redshift ranges for various survey sensitivities using the counts from the figure (4). EMU-ASKAP and WODAN are expected to have a sensitivity of 10uJy/beam. However, 5-sigma or more will be a more likely limit for science goals like EMU-Cosmology or similar studies using WODAN. Figure (7) shows shot-noise estimates for various power-law distributions ([42], [51]). Due to higher sensitivity, both EMU and WODAN are expected to get lesser shot-noise than VLASS as EMU and WODAN will have higher density per square degrees or steradians than VLASS ( 290 sources per square degtrees)[47], [48] .

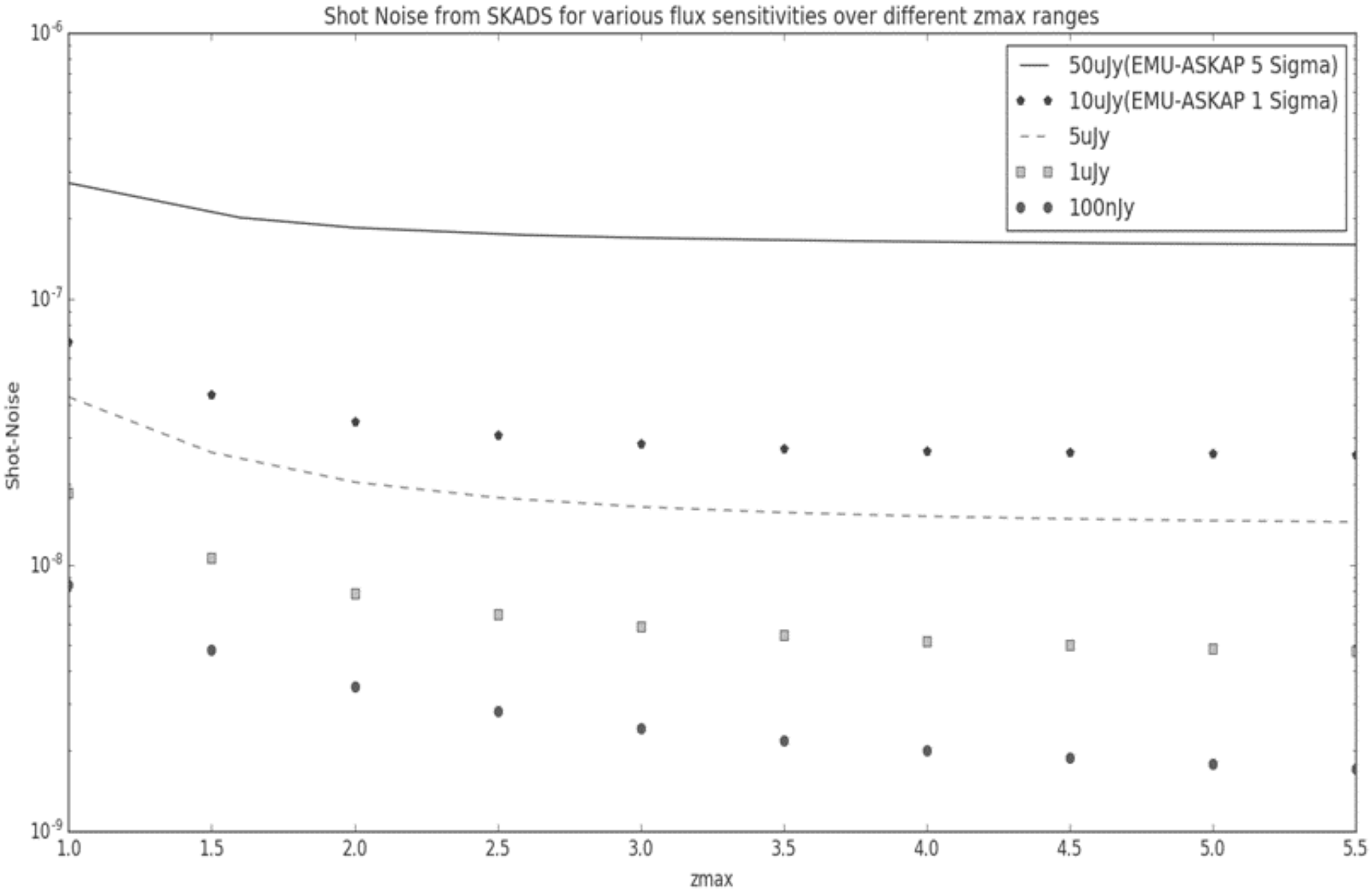

**Figure 6-Shot-noise estimates over different redshift ranges for various survey sensitivities. EMU-ASKAP and WODAN are expected to have a sensitivity of 10uJy. However, 5-sigma or more will be a more likely limit for science goals like EMU-Cosmology (Norris et al. 2011) or similar studies using WODAN (54). A more sensitive survey like MIGHTEE tier-2 (Jarvis 2012, Hales 2013) will go as far as 1uJy/beam and so will result in a much lesser shot-noise.**

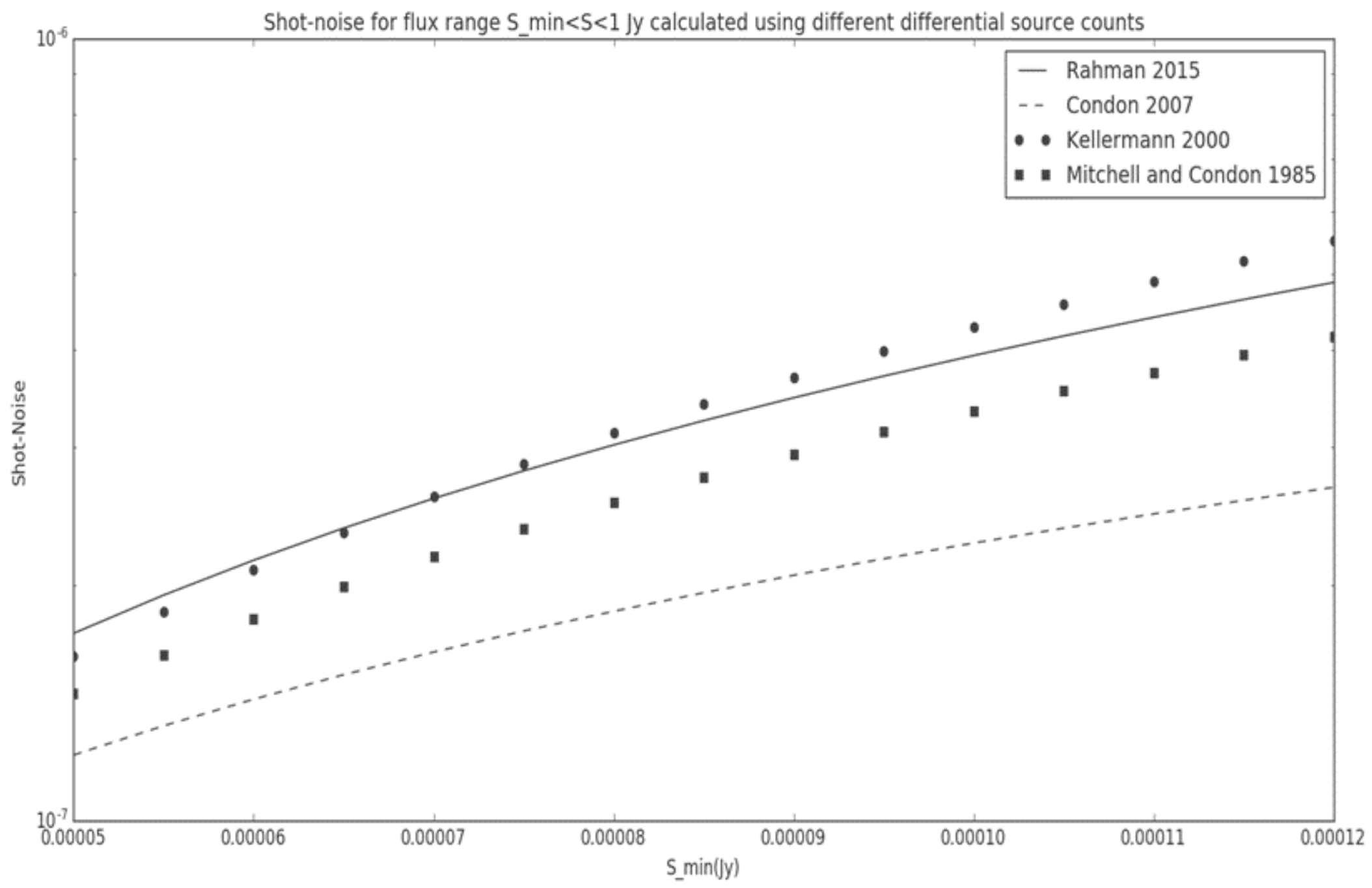

**Figure 7-Shot-Noise estimates using the power law distributions given in Rahman 2015, Condon 2007, Kellermann 2000 and Mitchell and Condon 1985. Here we restricted Smax=1Jy/beam.**

## SIGNAL TO NOISE RATIOS FOR ISW ESTIMATES

A major scientific goal of modern radio continuum surveys and the cosmic microwave background surveys is the measurement of the late time integrated Sachs-Wolfe (ISW) effect ([26], [27]). The effect explains the blue-shifting of the photons from the cosmic microwave background rations once they pass through the large scale structures or the red-shifting of the photons once they pass through big voids. The detection of ISW effect is used as a mean to constraint the dark energy ([26], [27],[57],[58],[59], and [60]) which explains the accelerated expansion of our universe ([13],[15],[16],[17],[20] and [60]), the age of our universe ([58], and [59]) and various other interesting phenomena.

The temperature fluctuations, due to the ISW effect, can be calculated using the gravitational potential as (read ref: [18], [27]):

$$\frac{\Delta T}{T} = \int_{\eta r}^{\eta 0} (\Phi' - \Psi') d\eta$$

Where, $\Phi = -\Psi$, in the linear regime for the conformal Newtonian gauge.

However, in fact, we measure a coefficient 'Cl' ([9], [18]) by cross-correlating the cosmic microwave background anisotropy maps with galaxy over/under density maps.

The cross-correlation angular power spectrum coefficient 'Cl' can be calculated as ([9], [18]):

$$Cl^{gt} = 4\pi \int_{kmin}^{kmax} \frac{dk}{k} \Delta^2(k) Wl^g(k) Wl^t(k) \qquad (9)$$

For more detailed discussions on cross-correlation angular power spectrum coefficient, please read ref: [28, 47-52].

For our analysis, we will use limber approximations [61]. We use CAMB [62] to obtain the matter power spectrum values. For our analysis, we adopt $\sigma_n = 50\mu Jy/\mathrm{beam}$ ([8], [9]).

Error bars for the 'Cl' values in equation (9) can be calculated as ([9], [18], and [52]):

$$\Delta Cl^{gt} = \sqrt{\frac{((Cl^{gg} + Shot{-}Noise)Cl^{tt} + (Cl^{gt})^2)}{(2l+1)fsky}} \qquad (10)$$

Here, Clgg is the galaxy-galaxy auto-correlation angular power spectrum, Cltt is the total CMB power spectrum, 'l' is the multipole value for the coefficient 'Cl' is being calculated and fsky is the ratio of the sky covered in the study.

Using equations (9 and 10), the signal to noise ratio (SNR) can finally be calculated as ([9], [52]):

$$(SNR)^2 = \sum_l \left(\frac{Cl^{gt}}{\Delta Cl^{gt}}\right)^2 \quad (11)$$

From, equations (9, 10 and 11), we can observe that the signal to noise measurements are dependent on the survey redshift range and the multipole ranges used in the analysis.

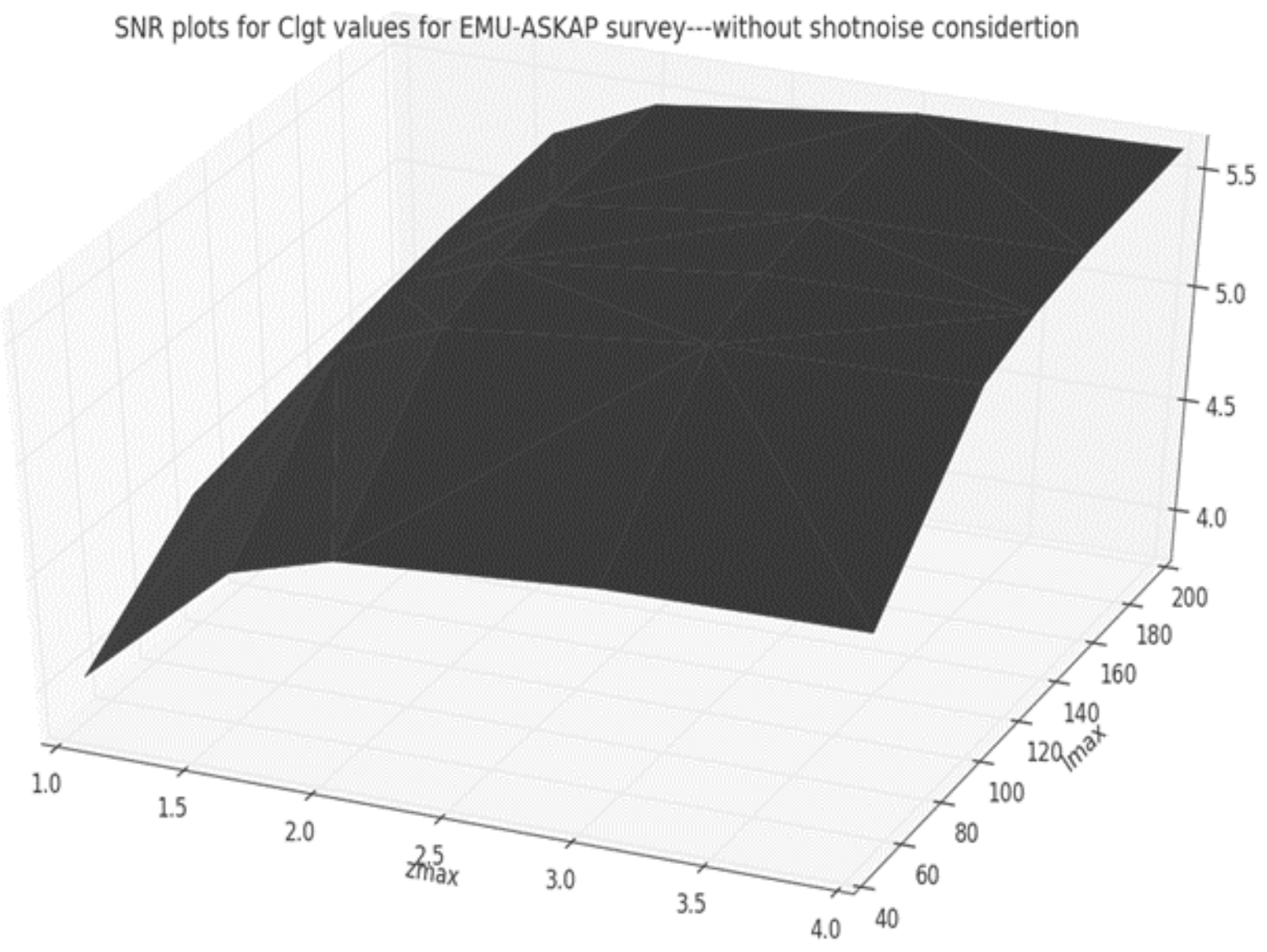

**Figure 8- Signal to noise ratio estimates for EMU-5Sigma like surveys with fsky=0.5. The signal-to-noise ratios are taken for various maximum redshift ranges and maximum multipole ranges.**

Figure (8) estimates the signal to noise ratios for ISW effect using EMU-5 sigma like surveys with ideal survey conditions i.e. no shot-noise. However, in order to get some more accurate estimates, we need to consider shot-noise as part of our analysis. Figure (9) shows the signal to noise ratio plots with shot-noise consideration and figure (10) shows the difference in SNR estimates for without and with shot-noise considerations.

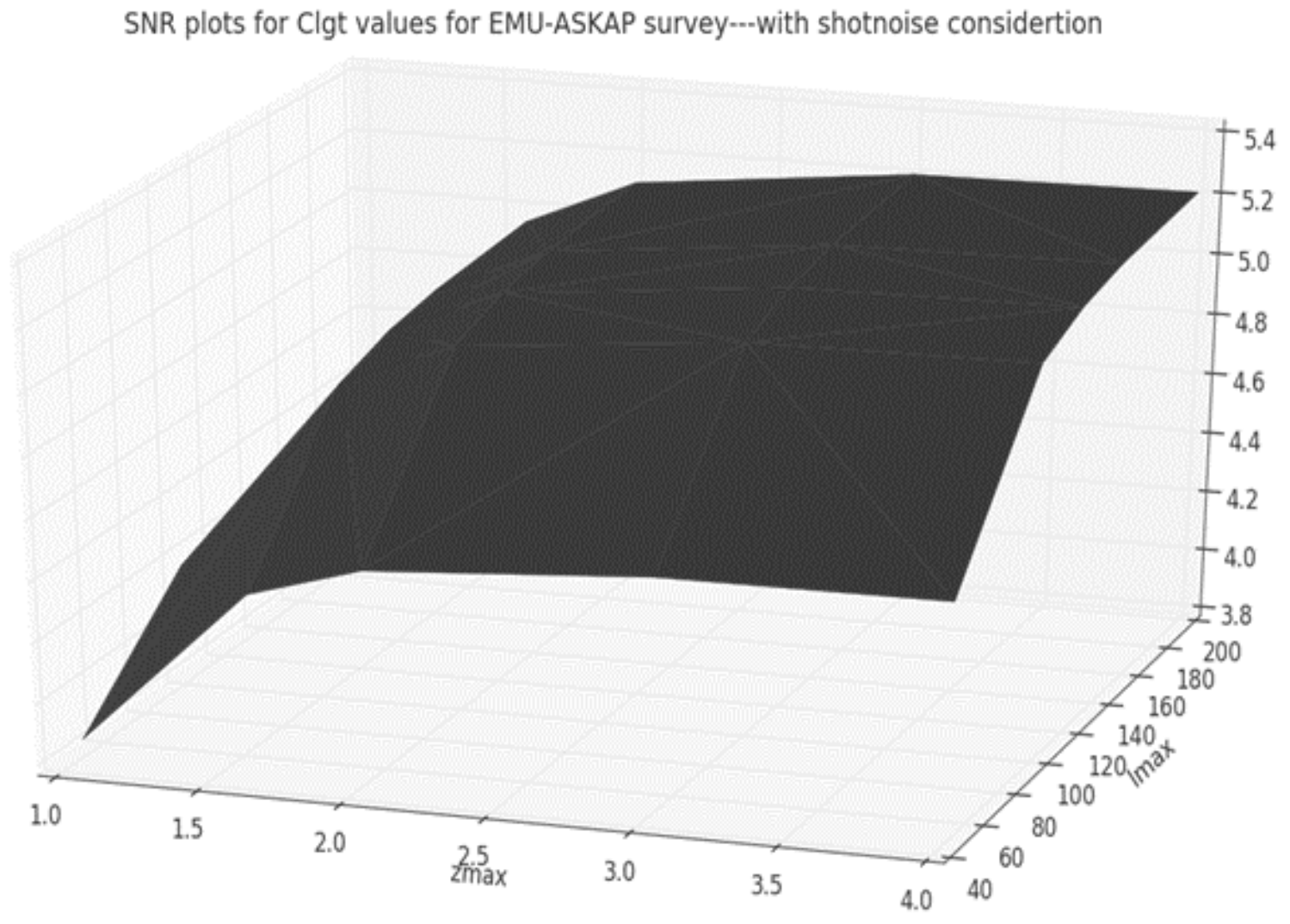

**Figure 9-Signal to noise ratio plots over different maximum redshift and 'l' ranges using fsky=0.5 and considering shot-noise for EMU-5Sigma like surveys**

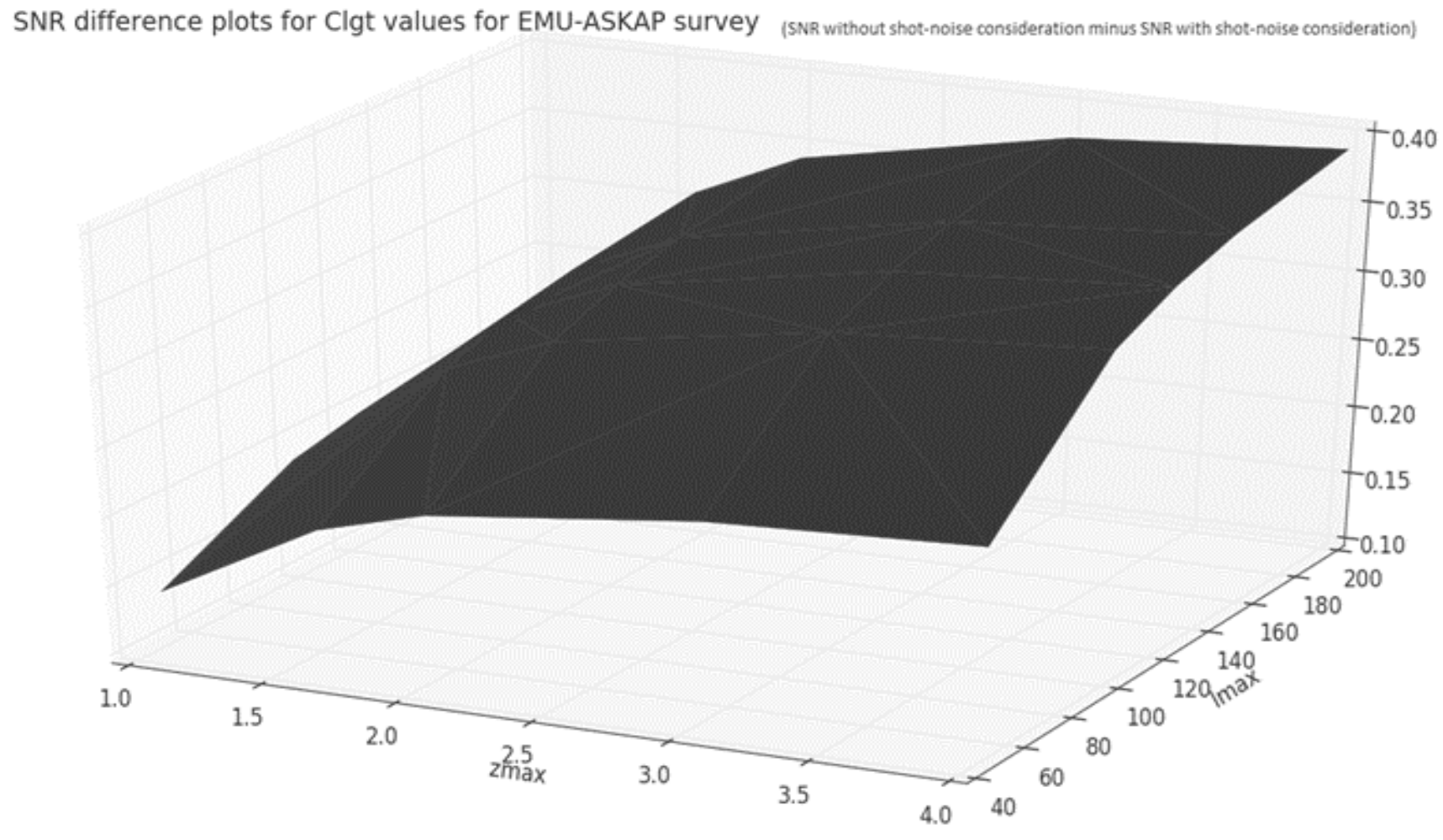

**Figure 10-SNR difference plots for EMU-5 sigma like surveys. The plot shows the difference in SNRs between without shot-noise consideration and with shot-noise consideration cases.**

# MAPPING ISSUES

Mapping the over/under density is a crucial part of the observational analysis in galaxy-galaxy auto-correlations or galaxy-CMB cross-correlations to observe an effect like the late time integrated Sachs-Wolfe Effect.

Surveys like Planck ([14],[16]) and WMAP ([10], [63]) have provided high-resolution maps for temperature fluctuations in the cosmic microwave background radiations. The maps are in the HEALPix[1] format ([64], [65]) which is a standard format for such studies. In order to develop compatible galaxy over/density maps, we need to adopt various strategies.

In a usual process, we first need to identify the galactic plane regions and regions with the unwanted signal. These foreground areas will be masked and excluded from the analysis.

Then we can take an average galaxy count as:

AC= N/P

Where,

AC= Average count per pixel

N = Total galaxies in the number count map (not the catalog as it will have masked objects too)

P = Total number of pixels in the area of the useful map we are interested in.

The total number of pixels in the area of the map, we are interested in, can be calculated using the mask (or the combined mask of the foreground and the area not covered in the survey). Use the masked map to calculate total pixels in the map as there will be 0s in the original catalog maps too, which represents the genuine absence of an object in the given map. A masked pixel can be assigned the value of 0 and the valid pixel is assigned the value of 1. Then we get the total count of '1' value pixels and which will give us 'total pixels in the area of the map'.

[1] http://healpix.sourceforge.net

Now, to get total galaxies in the number count map, we need to make sure that objects in the masked region are not counted.

We will get right ascension and declination or other positions in other coordinate systems for each object from the catalog. We will then have to convert them into the HEALPix map pixel format. If the pixel lies in the masked region, then we can exclude it from the count. By this method, we can count all the objects in the catalog which are not in the masked region. This will give us 'total galaxies in the galaxy number count map'. Once we have got total pixels and total galaxies, then we can get the average count per pixel as discussed before.

After this, we will need to develop a count per pixel map which is simply achieved by converting all object positions into their HEALPix counterparts and then we count the repeating positions. This will give us the number of galaxies per pixel in the survey. Now comes the over/under density part. This can be calculated as:

$$\mathrm{ODP}=\frac{\mathrm{NCP}-\mathrm{AC}}{\mathrm{AC}}$$

Where,

NCP=Galaxy count per pixel

ODP= Over/Under density per pixel

This can be done by creating a function which will take the galaxy number count per pixel map and average count per pixel map. Then we can iterate through each pixel in the map and use the over/under density formula.

Once we got the over/under density map, then we can use the combined mask and this over/under density map with a utility such as HEALPix's ([64], [65]) 'anafast' function or similar methods to get the Cl values. 'Cl ' values are basically derived from another coefficient 'alm', usng:

$Cl=<|a_{lm}|^2>$

'alms' are the real measure of deviations and their expected value for a smooth sky is '0'. For the Cl calculations, their absolute value is used means both under and over densities increase alms and so Cls. Unusually high source count in some regions will increase over densities in some regions and under densities in other regions which will result in some higher Cls.

In NVSS, there were a few major issues like object over-lapping and position uncertainty based on declination ([22], [29]). Since EMU-ASKAP will have a much better depth and resolution, so we can expect better results in this regard too. We will just need to properly identify a suitable map making strategy based on our science goals.

We will first need to identify the target resolution we need, based on position accuracy, side lobe error removal, weak source and maximum multipole (l) range required for the science goal. Then, based on that target resolution, we can decide the NSide value for HEALPix maps. NSide value determines the resolution ([64], [65]).

Maximum multipole (l) should optimally not exceed 2*Nside. EMU will have a lot more depth than NVSS ([1], [35], [36]), sensitivity and higher resolution so we can go for a much optimum solution even during the map making stage. We can go for higher resolution maps like NSide =1024 or more. This will help us in avoiding the smoothing or other post image approximations. This will also reduce the requirement for using any other workaround or arbitrary offsets.

A possible way, to avoid such errors at map making stage, is:

- We can assume a target map resolution, which can be based on our survey resolution, rms confusion limit, and position accuracy. Then we can choose a very high Nside value to get mean pixel area equal to the square of our target resolution (square root of the pixel area and then converted to arcsec).
- In the next step, pixels with 0 sources should be considered as 0 and values more than 1 should be considered as 1. This means if a pixel has 2, 3 or more sources then it should be considered as 1 so that we can avoid side lobe errors or non-galactic sources.
- Once, we are sure that we have removed most of the unwanted sources, then we can reduce the map's resolution by either re-developing the map using the high-resolution count map or by simply reducing the map obtained in the previous step and simply apply a method like the HEALPix's ([64], [65]) **ud_grade.**
- This will also determine how big our maps will be and how much processing power we will need to process the data.

## SKY COVERAGE AND SIGNAL TO NOISE RATIOS

The masking strategies and flux-cut off ranges discussed in previous sections, come with their own trade-offs. As we can see in equations (10) and (11) that the signal to noise ratio is also dependent on the sky coverage ratio (fsky) and also on the flux range or galaxy counts via shot-noise.

In figure (11), we presented a comparison of EMU-ASKAP with WODAN, EMU-ASKAP+WODAN and EMU-Early Science. EMU-ASKAP will have a sky coverage ratio fsky ≈ 0.75 but with foreground masking ([9], [14], [16]) due to the galactic plane, solar system objects, and other contaminated areas, it can be expected to go down to about fsky ≈ 0.5. However, we can use the compatibility of WODAN survey with EMU-ASKAP survey to increase the overall useful sky coverage for measuring the late time ISW effect to about fsky ≈ 0.75. EMU-Early Science Cosmology survey is likely to utilize 2000 square degrees of the sky [43]. In figure (11), we plotted results without shot-noise consideration. In figure (12), we can see results with shot-noise.

EMU-Early Science on its own will not give us any significant cosmological results when it comes to detecting the late time integrated Sachs-Wolfe effect. However, it will help us in testing the development pipeline, understanding the expected ASKAP performance parameters and may also be helpful in improving our statistical redshift estimates, especially when cross-matched with optical survey catalogs of the southern sky.

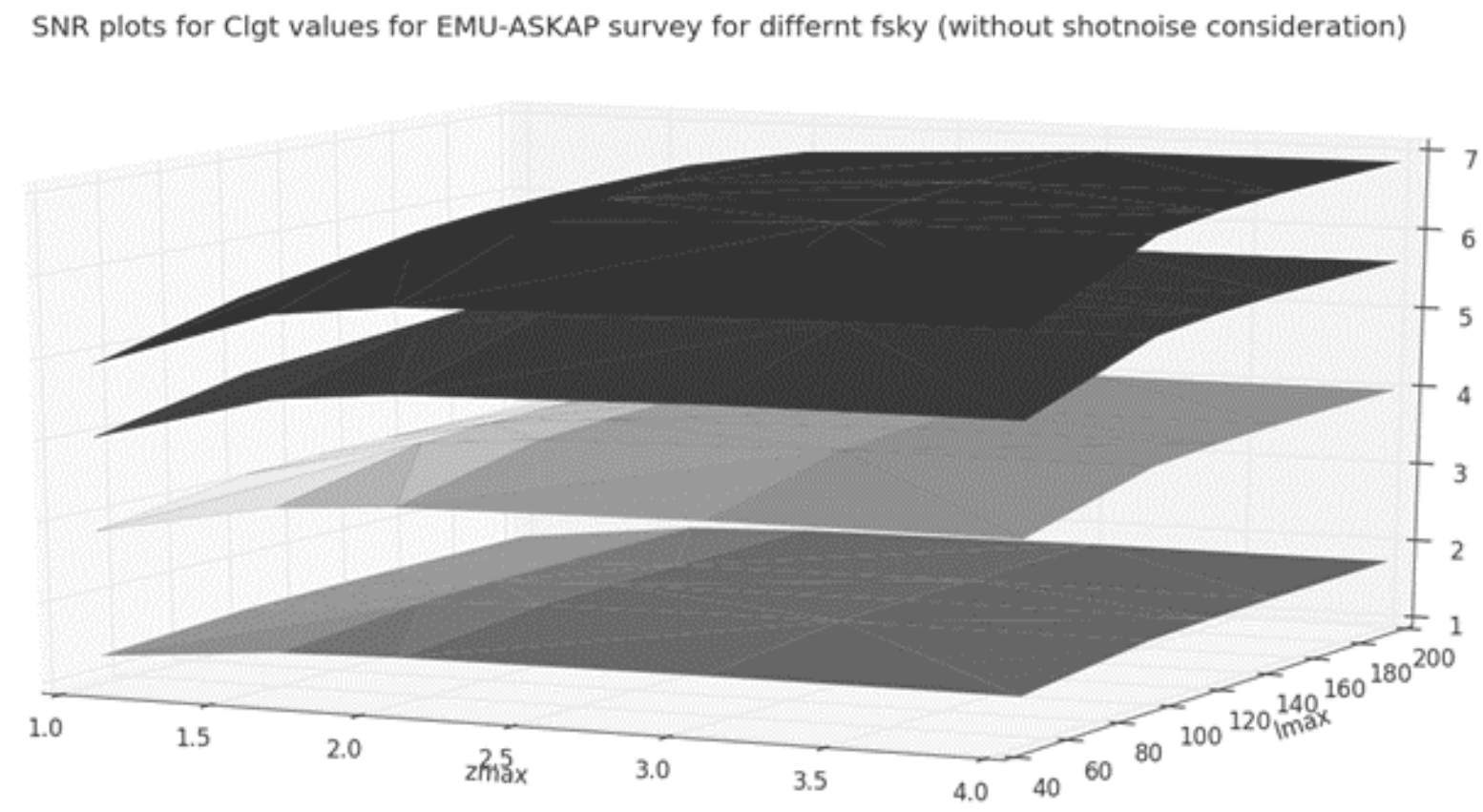

**Figure 11-Signal to noise ratio plots for EMU-ASKAP full with fsky=0.5 (second tier). Also shown are plots for EMU-ASKAP+WODAN (top tier), WODAN alone (third tier) and EMU-Early Science (bottom). Here we did not include shot-noise in our calculations.**

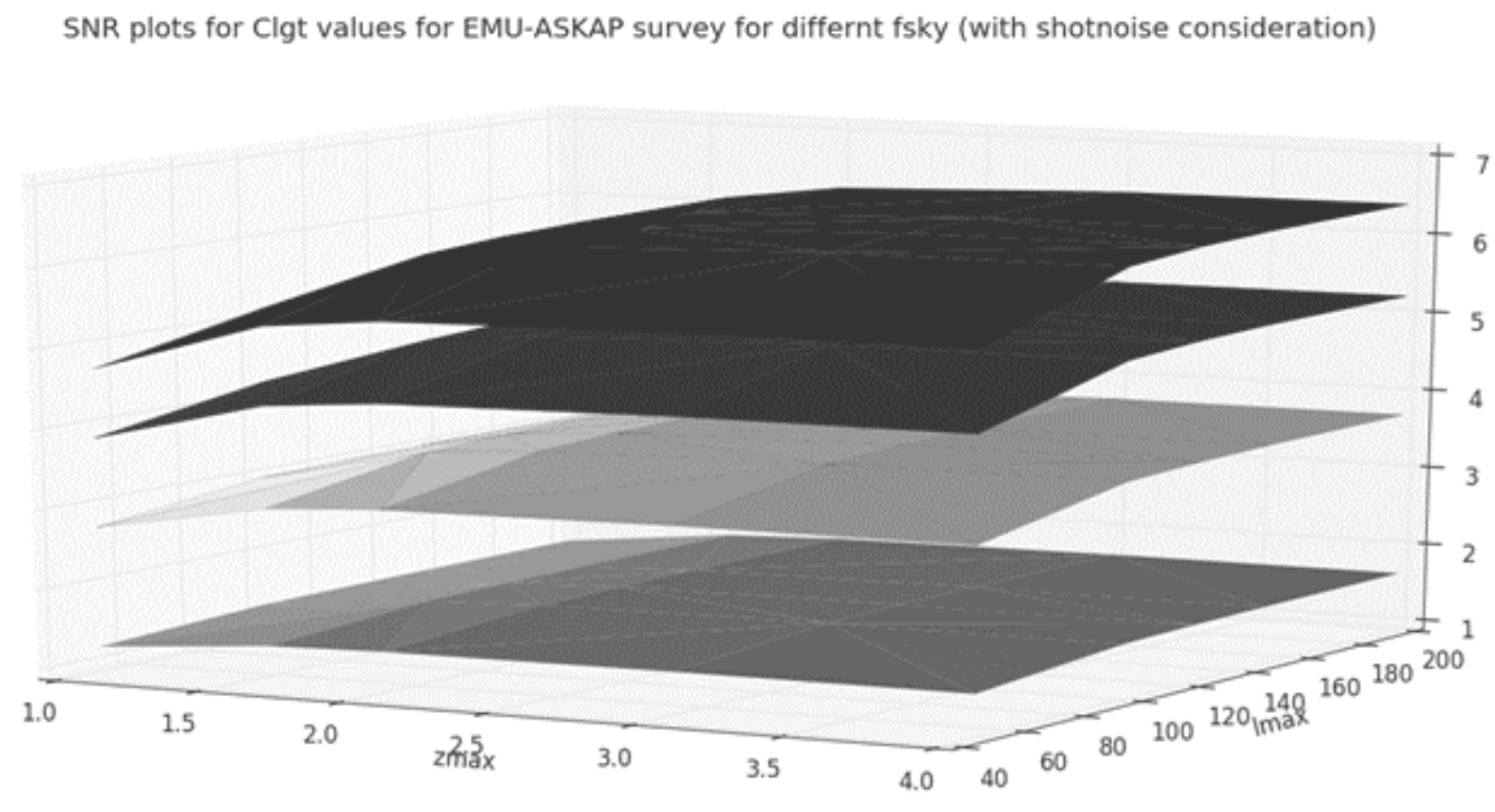

**Figure 12-Signal to noise ratio plots for EMU-ASKAP full with fsky=0.5 (second tier). Also shown are plots for EMU-ASKAP+WODAN (top tier), WODAN alone (third tier) and EMU-Early Science (bottom). Here we included shot-noise in our calculations.**

From figures (8,9,10,11 & 12), we can observe the effects of power-law source counts and shot-noise, as discussed in earlier sections, on the significance of an important science goal like the late time integrated Sachs-Wolfe effect.

We can observe in figure (12) that a combined EMU and WODAN galaxy over/under-density map will increase the signal to noise ratios significantly for different redshift and lmax ranges. EMU and WODAN will complement each other in cosmological studies, like ISW effect, galaxy auto-correlations and cosmic magnification due to their similar sensitivity and their coverage of different regions of the sky as discussed earlier. VLASS is expected to have comparable sky coverage as EMU but due to lesser sensitivity will have lower S/N ratios for science goals like the ISW effect. However, VLASS will have better S/N ratio in comparison with WODAN due to almost three times better sky coverage despite having lower sensitivity.

## CONCLUSION

In this study, we observed how issues like source counts, shot-noise, confusion, position accuracy, mapping issues and sky coverage affect our science goal results. We also observed power-law distributions can affect number counts per steradian estimates, shot-noise estimated and rms confusion calculations. We particularly discussed the expected performance of upcoming surveys like EMU-Early Science, EMU-ASKAP full and WODAN. The discussion and results will also be useful for future continuum surveys like with possible continuum surveys through VLASS, MEERKAT (e.g. MIGHTEE) and SKA-1. We found that in order to better understand the significance of our observational results, we not only need to focus on the scientific aspects of our observations and related data but also the technical assumptions we are making regarding the instruments we are using. We also discussed the impact of masking issues and survey coverage on the significance of science goals and based on the theoretical estimates, also concluded that with combined EMU and WODAN data analysis, we can achieve significant improvements in signal to noise ratios for science goals like ISW effect. EMU and WODAN are highly compatible surveys based on their flux sensitivity and their coverage of different regions of the sky.

## ACKNOWLEDGMENTS

I would like to thank Prof. Dr. Jeremy Mould from Centre for Astrophysics and Supercomputing at the Swinburne University of Technology for reading and providing valuable suggestions to improve the quality of this paper.

I would also like to thank all my colleagues and friends who pointed out typos and asked useful questions which helped in improving the quality of this paper. Some of the results in this paper have been derived using the HEALPix (K.M. Górski et al., 2005, ApJ, 622, p759) package ([64], [65]).